\documentclass[12pt]{iopart}

\usepackage{iopams}
\usepackage[dvipsnames,rgb,dvips]{xcolor}
\usepackage{overpic}
\usepackage{graphicx}
\usepackage{dcolumn}
\usepackage{epstopdf}
\usepackage{siunitx}
\usepackage{mathrsfs}
\makeatletter
\def\amsbb{\use@mathgroup \M@U \symAMSb}
\makeatother
\usepackage{bbm}
\expandafter\let\csname equation*\endcsname\relax
\expandafter\let\csname endequation*\endcsname\relax
\usepackage{amsmath}

\newcommand{\ve}[1]{\boldsymbol{#1}}
\newcommand{\ma}[1]{\ensuremath{\amsbb{#1}}}

\newcommand{\gghat}{\ensuremath{\hat{\ve g}}}

\newcommand{\rd}{{\rm d}}

\newcommand{\eqnlab}[1]{\label{eq:#1}}
\newcommand{\figlab}[1]{\label{fig:#1}}
\newcommand{\seclab}[1]{\label{sec:#1}}
\newcommand{\eqnref}[1]{(\ref{eq:#1})}

\newcommand{\Eqnref}[1]{Eq.~(\ref{eq:#1})}
\newcommand{\Figref}[1]{Fig.~\ref{fig:#1}}

\newcommand\etaK{\eta_{\rm K}}
\newcommand\tauK{\tau_{\rm K}}
\newcommand\tauL{\tau_{\rm s}}
\renewcommand{\tr}{{\rm Tr}}
\newcommand{\ku}{{\rm Ku}}
\newcommand{\st}{{\rm St}}
\newcommand{\sv}{{\rm Sv}}
\newcommand{\re}{{\rm Re}}
\newcommand{\St}{{\rm St}}
\newcommand{\rep}{{\re_{p}}}

\def\onedot{$\mathsurround0pt\ldotp$}
\def\cddot{% two dots stacked vertically
  \mathbin{\vcenter{\baselineskip.67ex
    \hbox{\onedot}\hbox{\onedot}}%
     }}%

\begin{document}
\title[Orientation of a small spheroid settling in turbulence]{Theory for the effect of fluid inertia on the orientation of a small spheroid settling in turbulence}
\author{K.~Gustavsson$^1$, M.~Z.~Sheikh$^2$, D.~Lopez$^{3}$, A.~Naso$^3$,  A.~Pumir$^2$, and B. Mehlig$^1$}
\address{$^1$ Department of Physics, Gothenburg University, 41296 Gothenburg, Sweden \\
$^2$ {Univ. Lyon, ENS de Lyon, Univ. Claude Bernard, CNRS, Laboratoire de Physique, F-69342,
Lyon, France}\\
$^3$  {Univ. Lyon, Ecole Centrale de Lyon, Univ. Claude Bernard, CNRS, INSA de Lyon, Laboratoire de M\'ecanique des Fluides et d'Acoustique, F-69134, Ecully, France}}

\begin{abstract}
Ice crystals settling through a turbulent cloud are rotated by turbulent velocity gradients.  In the same way, turbulence affects the orientation of aggregates of organic matter settling in the ocean.  In fact most solid particles encountered in Nature are not spherical, and their orientation affects their settling speed, as well as collision rates between particles.  Therefore it is important to understand the distribution of orientations of non-spherical particles settling in turbulence. Here we study the angular dynamics of small prolate spheroids settling in homogeneous isotropic turbulence. We consider a limit of the problem where the fluid torque due to convective inertia dominates, so that rods settle essentially horizontally. Turbulence causes the orientation of the settling particles to fluctuate, and we calculate their orientation distribution for prolate spheroids with arbitrary aspect ratios for large settling number $\sv$ (a dimensionless measure of the settling speed), assuming small Stokes number $\st$ (a dimensionless measure of particle inertia). 
This overdamped theory predicts that the  orientation distribution is very narrow at large
$\sv$, with a variance proportional to  $\sv^{-4}$. By considering the role of particle inertia, we analyse the limitations of the overdamped theory, and determine its range of applicability. Our predictions are in excellent agreement with numerical simulations of simplified models of turbulent flows. Finally we contrast our results with those of an alternative theory predicting that the orientation variance scales as $\sv^{-2}$ at large $\sv$.
\end{abstract} 

\maketitle

\section{Introduction}
The settling of particles in turbulence is important in a wide range of scientific problems. An example is the settling of small ice crystals in turbulence, a process that is considered in the context of rain formation from cold cumulus clouds \cite{Pru78,Cho81,Hub14}. Further  examples are  the settling of small aggregates of organic matter (\lq marine snow\rq{})  \cite{Kio01}, and the dynamics of swimming microorganisms   \cite{Rui2004,Cen13,Berglund2016} in the turbulent ocean.

The settling of spherical particles in turbulence has been intensively studied. Maxey and collaborators \cite{MC86,Max87,Wan93} found that turbulence increases the settling speed of  small spherical particles. This pioneering work has led to many experimental and numerical studies, using direct numerical simulation (DNS) of turbulence, and it is a question of substantial current interest \cite{good_ireland_bewley_bodenschatz_collins_warhaft_2014,petersen_baker_coletti_2019}. An important question is how frequently particles collide as they settle in turbulence \cite{Ard16,For19}. The collision rate is influenced by spatial inhomogeneities in the particle-number density due to the effect of particle inertia. There is substantial recent progress in understanding this two-particle problem \cite{Gus14e,Bec14,Ireland,Mathai2016,Parishani}. The conclusion is that settling may increase or decrease spatial clustering of spherical particles, and that it tends to decrease the relative velocities of nearby particles because settling reduces the frequencies of \lq caustics\rq{},
singularities in the inertial-particle dynamics \cite{Gus14e}.

Most solid particles encountered in Nature and in Engineering are not spherical, yet less is known about the settling of non-spherical particles in turbulence, and their settling depends in an essential way on their orientation.
In a fluid at rest the orientation of a slowly settling non-spherical particle is determined by weak torques induced by the  convective inertia of the fluid - set in motion by the moving particle.
For a single, isolated particle in a quiescent fluid this effect is well understood \cite{Cox65,Kha89,Dab15,Can16}: convective fluid inertia due to slip between the particle and the fluid velocity causes non-spherical particles to settle with their broad side first.  For axisymmetric rods, for example, symmetry dictates that the angular dynamics has two fixed orientations: either the rod is aligned with gravity (tip first) or  perpendicular to gravity. At weak inertia, only the latter orientation is stable, so that the rod settles with its  long edge first. But when there is turbulence, then turbulent vorticity and strain exert additional torques that cause fluctuations in the orientations of the settling crystals \cite{Pru78,Kle95}.

To understand the angular motion of  a non-spherical particle settling in turbulence is in general  a very complex problem, because there are many dimensionless parameters to consider.  There is particle shape (shape parameter $\Lambda)$, and the effect of particle inertia is measured by the Stokes number $\st$. The importance of settling is determined by $\sv$, a dimensionless measure of the settling speed. The significance of fluid inertia is quantified by two Reynolds numbers, the particle Reynolds number $\re_p$ (convective inertia due to slip between particle and fluid velocity), and the shear Reynolds number $\re_s$ (convective inertia due to fluid-velocity gradients). The nature
of the turbulent velocity fluctuations is determined by the Taylor-scale Reynolds number $\re_\lambda$.

If the particles are so small that they just follow the flow and that any inertial corrections to the fluid torque are negligible ($\re_p=\re_s=0$), then the angular dynamics of small crystals in turbulence
is well understood \cite{Jef22,Pum11,Par12,Che13,Gus14,Byr15,Zha15,Voth15,Voth16,Berglund2016,Fri17}. The particle orientation responds to local vorticity and strain through Jeffery's equation \cite{Jef22}. The effect of particle inertia is straightforward to take into account \cite{einarsson2014}, but the role of fluid inertia is more difficult to describe, even in the absence of settling.
In certain limiting cases fluid-inertial effects are well understood. The most important example is that of a small neutrally buoyant ($\re_s=\st$)
spheroid moving in a time-independent linear shear flow, so that the centre-of-mass of the particle follows the flow ($\re_p=0$).  Neglecting inertial effects ($\re_s=0$) and angular diffusion, the angular dynamics degenerates into a one-parameter family of marginally stable orbits, the so-called Jeffery orbits \cite{Jef22}. Fluid inertia breaks this degeneracy and gives rise to certain stable orbits  \cite{saffman1956,subramanian2005,einarsson2015a,rosen2015d}. Much less is known when $\re_p$ is not zero. Candelier, Mehlig \& Magnaudet \cite{Candelier2018} recently showed how to  compute  the effect of  a small slip  upon the force and torque on a non-spherical particle in a general linear time-independent  flow, by generalising
Saffman's result \cite{Saf65,Saf68} on the lift upon a small sphere in a shear flow, valid in the limit where $\re_p\ll\sqrt{\re_s}\ll 1$.

The results summarised in the previous paragraph pertain to time-independent flows. Time-dependent spatially inhomogeneous flows present new challenges, and very little is known about the effect of fluid inertia for such flows, in particular for turbulence. In some studies, therefore, effects of fluid inertia were simply neglected \cite{Siew14a,Siew14b,Gus17,Jucha2018,Naso2018}. These models predict that the breaking of isotropy due to gravity causes a bias in the orientation distribution
of the settling particles, so that rods tend to settle tip first, parallel to gravity.  For small particles it is safe to neglect $\re_s$ \cite{Candelier2016}.  But experiments and numerical simulations of slender particles settling in a  vortex flow \cite{Lop17} and in turbulence \cite{Koc16} show that  convective inertial  torques due to settling can
make a qualitative difference to the orientation distribution.

In this paper we therefore consider the effect of the convective inertial torques on the orientation of small spheroids settling in turbulence.  Following Ref.~\cite{Lop17}, our model assumes that the hydrodynamic torque is approximately given by the sum of Jeffery's torque and the convective inertial torque in a homogeneous, time-independent flow.
For nearly spherical particles this convective torque was calculated
by Cox \cite{Cox65}, and for slender bodies  by Khayat \& Cox \cite{Kha89}. Their results were generalised to spheroids with arbitrary aspect ratios in Ref.~\cite{Dab15}.

Our goal is to analyse how the turbulent-velocity fluctuations affect the orientation distribution of a prolate spheroid settling through turbulence. We assume that the particles are small enough so that convective-inertia effects due to the fluid-velocity gradients are negligible, that inertial effects
on the centre-of-mass motion are small (small $\st$ and $\re_p$), but that the settling number $\sv$  is large enough so that the fluid-inertia torque dominates the angular dynamics.

We find an approximate theory for the angular
distribution of settling spheroids using a statistical model \cite{Gus16,Gus17} for
the turbulent fluctuations. The theory is valid for large $\sv$ and small  $\st$, in the overdamped limit, and its predictions are in excellent agreement with results of numerical simulations of the statistical model, and with simulations using a kinematic-simulation (KS) model \cite{Fung:92,Vosskuhle:15} for the turbulent flow. We find that the variance
of the orientation scales as $\sv^{-4}$ in the limit of  large settling number $\sv$,
for small enough Stokes number $\st$, and the theory determines how the pre-factor depends on the
shape of the spheroid. In the slender-body limit, the $\sv^{-4}$-scaling of the variance was also found in Ref.~\cite{Kramel} using an approach equivalent to ours.

We contrast our results with a theory for the orientation variance derived by Klett \cite{Kle95} for nearly spherical particles. This theory predicts that the variance is proportional to $\sv^{-2}$. At first sight this  may appear to be at variance with the overdamped theory, but
we show that  the overdamped  approximation breaks down into several different regimes when particle inertia begins to matter.
At very large values of $\sv$,  when the time scale at which the fluid-velocity gradients decorrelate is the smallest time
scale of the inertial dynamics, our numerical simulations show a $\sv^{-2}$-scaling, as suggested by Klett's theory. 
But the theory is difficult to justify because it neglects particle inertia in the centre-of-mass dynamics.
We show that translational particle inertia has a significant effect upon the angular dynamics, so that
it must be taken into account as soon as the overdamped approximation breaks down.
 
The remainder of this paper is organised as follows. In Section \ref{sec:formulation} we describe our model: the approximate equations of motion and the statistical model for the turbulent-velocity fluctuations. In Section
\ref{sec:od} we show results of numerical simulations of our model. We describe how
and why the results differ from those in Refs.~\cite{Siew14a,Siew14b,Gus17,Jucha2018,Naso2018},
and explain the intuition behind our theory for small $\st$ and large $\sv$. The overdamped theory is
described in Section \ref{sec:advective}.  Section \ref{sec:underdamped} discusses the effect of particle
inertia, and Section \ref{sec:conclusions} contains our conclusions as well as an outlook.

\section{Model}
\label{sec:formulation}
\subsection{Particle equation of motion}
Newton's equations of motion for a single non-spherical particle read:
\begin{subequations}
\label{eq:eom}
\begin{eqnarray}
\label{eq:eomt}
&m_{\rm p} \dot{\ve v}_{\rm p} = \ve f + m_{\rm p} \ve g\,,\quad&\dot{\ve x}_{\rm p} = \ve v_{\rm p}\,,\\
&m_{\rm p}
\tfrac{\rd}{\rd t}
\big[\ma I_{\rm p}(\ve n) \ve \omega_{\rm p}\big]  = \ve \tau \,,\quad\,\quad\quad\quad &\dot {\ve n}= \ve \omega_{\rm p} \wedge \ve n\,.
 \end{eqnarray}
\end{subequations}
 Here $\ve g$ is the gravitational acceleration with direction $\gghat=\ve g/|\ve g|$, $\ve x_{\rm p}$ is the
position of the particle, $\ve v_{\rm p}$ its centre-of-mass velocity,
 $m_{\rm p}$ the particle mass, and the dots denote time derivatives.  We assume that the particle is axisymmetric, so that its orientation is characterised by the unit vector
$\ve n$ along the symmetry axis of the particle. The angular velocity
of the particle is denoted by  $\ve \omega_{\rm p}$,
and $\ma I_{\rm p}(\ve n)$ is its rotational inertia tensor per unit-mass in the lab frame. For a spheroid, the elements of $\ma I_{\rm p}(\ve n)$ are given by~\cite{Kim:2005}
\begin{eqnarray}
\label{eq:Ip}
(\ma I_{\rm p})_{ij}(\ve n)=I_\perp (\delta_{ij}-n_i n_j) + I_{\parallel} n_i n_j
\,,\quad
I_{\perp}=\frac{1+\lambda^2}{5}a_\perp^2
\,,\quad
I_{\parallel}=\frac{2}{5}a_\perp^2\,,
\end{eqnarray}
where $\lambda\equiv a_\parallel/a_\perp$ is the aspect ratio of the spheroid, $2a_\parallel$ is the  length of the symmetry axis, and
$2a_\perp$ is the diameter of the spheroid. Prolate spheroids correspond to $\lambda >1$, whereas oblate spheroids have $\lambda < 1$.

The difficulty  lies in computing the hydrodynamic force $\ve f$ and torque $\ve \tau$ on the particle.
In the Stokes approximation, unsteady and convective inertial effects
are neglected.
In this creeping-flow limit \cite{Kim:2005},
the force  and   torque upon the  spheroid  are linearly related to the slip velocity $\ve W\equiv\ve v_{\rm p}-\ve u$,  to the angular slip velocity
$\ve \omega_{\rm p}-\ve \Omega$, and to the fluid strain $\ma S$:
 \begin{eqnarray}\label{eq:resistance}
\begin{bmatrix}
\ve f^{(0)} \\
\ve \tau^{(0)}
\end{bmatrix}
=
6\pi a_\perp\mu
\begin{bmatrix}
\ma A& 0 &0 \\
0 & \ma C & \ma H
\end{bmatrix}
\begin{bmatrix}
\ve u - \ve v_{\rm p} \\
\ve \Omega-\ve \omega_{\rm p} \\
\ma S
\end{bmatrix}\,.
\end{eqnarray}
Here $\mu$ is the dynamic viscosity of the fluid,  $\ve u\equiv\ve u(\ve x_{\rm p},t)$ is the undisturbed fluid velocity at the particle position $\ve x_{\rm p}$,
$\ve \Omega\equiv\frac{1}{2}\ve \nabla \wedge \ve u$
is half the vorticity of the undisturbed fluid-velocity field at the particle position,
and $\ma S$ is the strain-rate matrix,
the symmetric part of the matrix of the undisturbed fluid-velocity gradients
(its antisymmetric part is denoted by $\ma O$).
The tensors $\ma A$, $\ma C$, and $\ma H$ are translational and rotational resistance tensors.
Their forms are determined by the shape of the particle.
Eq.~(\ref{eq:resistance}) shows that the tensor $\ma A$ relates the hydrodynamic force $\ve f^{(0)}$
to the slip velocity $\ve W$. For an axisymmetric particle with fore-aft symmetry the tensor takes the form
\begin{equation}
A_{ij} \equiv
A_\perp (\delta_{ij}-n_i n_j) + A_\parallel n_i n_j\,.
\label{eq:def_Mt}
\end{equation}
The resistance coefficients $A_\perp$ and $A_\parallel$  depend on the shape of the particle.
For a spheroid, they are given by \cite{Kim:2005}:
\begin{eqnarray}
\eqnlab{eqm_C_coeffs}
A_\perp&=\frac{8(\lambda^2-1)}{3\lambda[(2\lambda^2-3)\beta+1]}\,,\;\;\;
A_\parallel=\frac{4(\lambda^2-1)}{3\lambda[(2\lambda^2-1)\beta-1]}\,,\\&\mbox{with}\;\;\;
\beta=\frac{\ln[\lambda + \sqrt{\lambda^2-1}]}{\lambda\sqrt{\lambda^2-1}}\,.
\nonumber
\end{eqnarray}
For a sphere one has $A_\perp=A_\parallel=1$, so that $\ve f^{(0)}$ simplifies
to the usual expression for  Stokes force on a sphere moving with velocity $\ve v_{\rm p}$ through a fluid
with velocity $\ve u$.

In the creeping-flow limit, the steady  slip velocity  $\ve W$  of a spheroid subject to a gravitational force $m_{\rm p} \ve g$ is obtained by setting the acceleration $\dot{\ve v}_{\rm p}$ to zero in Eq.~(\ref{eq:eomt}):
\begin{eqnarray}
\label{eq:Wslip}
\ve W^{(0)} &=\tau_{\rm p} \Big[ A_\perp^{-1}(\mathbbm{1}-\ve n\ve n^{\sf T})+ A_\parallel^{-1}\ve n\ve n^{\sf T}\Big] \ve g\,.
\end{eqnarray}
Here $\mathbbm{1}$ is the unit matrix, and $\tau_{\rm p}\equiv (2a_\parallel a_\perp\rho_{\rm p})/(9\nu\rho_{\rm f})$ is the particle response time in Stokes' approximation  with kinematic viscosity $\nu = \mu/\rho_{\rm f}$, fluid-mass density $\rho_{\rm f}$, and particle-mass density $\rho_{\rm p}$. The slip velocity depends on the orientation $\ve n$ of the particle.

For an axisymmetric particle with fore-aft symmetry, the rotational resistance tensors take the form:
\begin{eqnarray}
C_{ij} &\equiv
C_\perp (\delta_{ij}-n_i n_j) + C_\parallel n_i n_j
\quad\mbox{and}\quad H_{ijk} &=H_0\epsilon_{ijl} n_k n_l\,.
\label{eq:def_Mr}
\end{eqnarray}
Here $\epsilon_{ijl}$ is the  anti-symmetric Levi-Civita tensor, and we use the Einstein summation convention: repeated indices are summed from $1$ to $3$. For a spheroid, the rotational resistance coefficients read \cite{Kim:2005}:
\begin{eqnarray}
\eqnlab{eqm_K_coeffs}
C_\perp&=\frac{8a_\parallel a_\perp(\lambda^4-1)}{9\lambda^2 [(2\lambda^2-1)\beta-1]}\,,\;\;\;
C_\parallel&=-\frac{8a_\parallel a_\perp(\lambda^2-1)}{9(\beta-1)\lambda^2}\,,\\
H_0&= -C_{\perp}\frac{\lambda^2 - 1}{\lambda^2 + 1}\,.
\nonumber
\end{eqnarray}
Expressions (\ref{eq:resistance}) to \eqnref{eqm_K_coeffs} determine the hydrodynamic force and torque in the creeping-flow limit.
Fluid-inertia effects
are neglected in $\ve f^{({0})}$ and $\ve \tau^{({0})}$.

There are two distinct  corrections when fluid-inertia effects are weak but not negligible, due to the undisturbed fluid-velocity gradients, $\ma S$ and $\ma O$,
and due to the slip velocity $\ve W$.
 The former are parameterised by the shear Reynolds number $\re_s$, the latter by the particle Reynolds number $\re_p$:
\begin{eqnarray}
\label{eq:re_defs}
 \re_s &= \frac{s a^2}{\nu}\quad \mbox{and}\quad
  \re_p = \frac{W^{(0)}_\perp a}{\nu} \,.
\end{eqnarray}
Here  $a=\mbox{max}\{a_\perp,a_\parallel\}$ is the largest dimension of the particle, and
$W^{(0)}_\perp$ is an estimate of the slip velocity:
the magnitude of the velocity of a small slender spheroidal particle settling under gravity in a quiescent fluid with its symmetry axis perpendicular to gravity.  From Eq.~(\ref{eq:Wslip})
we see that $ W^{(0)}_\perp = \tau_{\rm p}g/A_\perp$. In the definition of $\re_s$, the parameter $s$ is a characteristic shear rate. In turbulence it is on average of the order
$s \sim\tauK^{-1}$ where $\tauK$ is the Kolmogorov time
\begin{eqnarray}
\label{eq:tauK}
\tauK &= {\big (2 \langle \tr \,\ma S \ma S^{\sf T}\rangle \big)^{-1/2}}\sim (\nu/\mathscr{E})^{1/2}\,.
\end{eqnarray}
Here  the average $\langle\cdots \rangle$
is over Lagrangian fluid trajectories, and $\mathscr{E}$ is the turbulent dissipation rate per unit mass.
This yields the estimate \cite{Candelier2016}
 $\re_s \sim (a/\etaK)^2$, where
 \begin{eqnarray}
\etaK &= \sqrt{\nu\tauK} \sim (\nu^3/\mathscr{E})^{1/4}
\end{eqnarray}
is the Kolmogorov length \cite{Fri97}. Thus the shear Reynolds number is small for small particles.

 Now consider the effect of convective inertia.
 Following Ref.~\cite{Lop17} we assume that the torque on the particle is given
 by the sum of Jeffery's torque and the instantaneous convective-inertia torque in a homogeneous flow.
 This approximation can be strictly justified for a steady linear flow  in the limit $\sqrt{\re_s} \ll \re_p \ll 1$. In this limit the singular perturbation problem that determines the fluid-inertia torque simplifies: the Saffman length ($\propto\re_s^{-1/2}$) is much larger than the Oseen length ($\propto {\re_p}^{-1}$). This implies that the leading convective-inertial corrections to the torque are those corresponding to a quiescent fluid, and a similar argument can be made for the convective-inertia contribution to the force. While there is no general theory explaining how the convective-inertia contributions to the force and the torque  are affected by spatial inhomogeneities in time-dependent flows, the results of Ref.~\cite{Lop17} show that the simple model used here can successfully explain important features of the orientation distribution of rods settling in a vortex flow.

\begin{figure}
\begin{overpic}[clip]{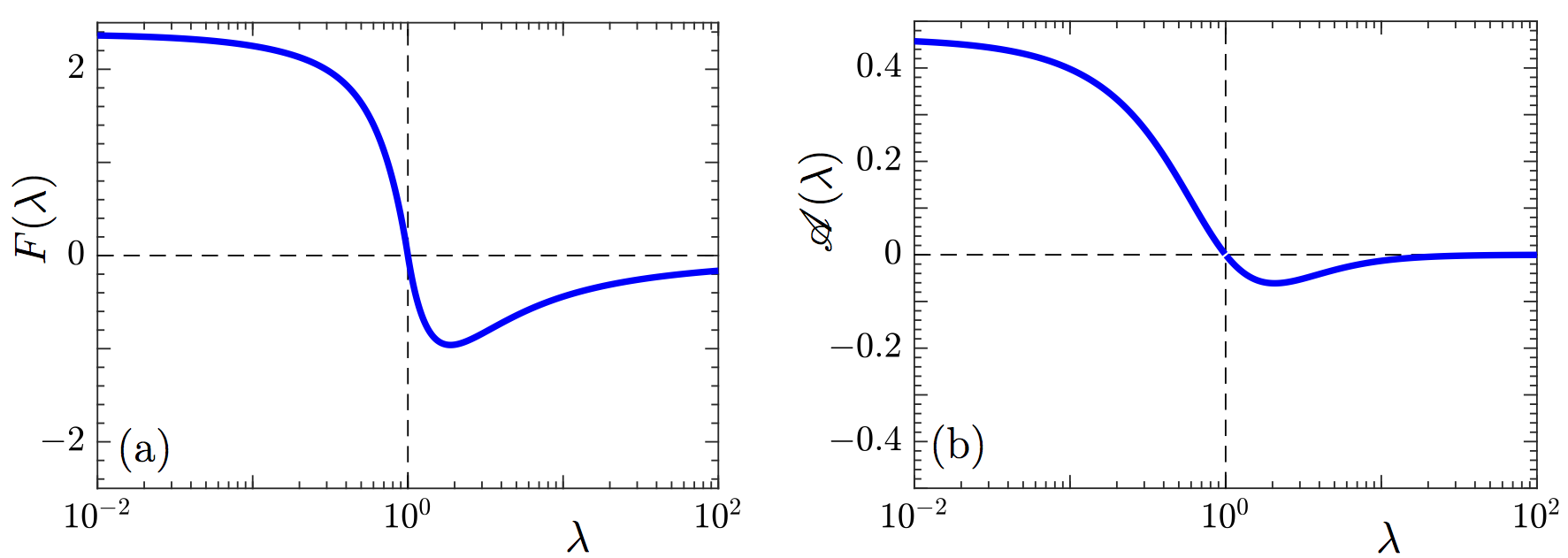}
\end{overpic}
\caption{\label{fig:shape_factor} Geometrical shape factors.
(a) Shape factor $F(\lambda)$ in Eq.~(\ref{eq:torque_fluid_inertia}).
The data shown are obtained by evaluating Eqs.~(4.1) and (4.2) in Ref.~\cite{Dab15}.
(b) Shape factor $\mathscr{A}(\lambda)$ defined in Eq.~(\ref{eq:defA}), as
a function of the particle aspect ratio $\lambda$.}
\end{figure}

 The leading-order inertial force correction reads  for a spheroid moving in a quiescent fluid \cite{Brenner61,Kha89}:
  \begin{equation}
 \label{eq:drag_correction}
\ve f^{(1)} = -(6\pi a_\perp\mu ) {\scriptstyle \tfrac{3}{16}}\re_p  \frac{W}{W_\perp^{(0)}}\big[3\ma A-\ma I (\hat{\ve W}\cdot \ma A \hat {\ve W})\big]\ma A \ve W\,,
 \end{equation}
with $W = |\ve W|$ and $\hat{\ve W} = \ve W/W$.
 For a spheroid,  the leading-order inertial contribution to the torque was calculated in Ref.~\cite{Dab15}:\begin{eqnarray}
\ve\tau^{(1)}= F(\lambda){\mu}   a^2\,\re_p \,\frac{W^2}{W_\perp^{(0)}}\,
 (\ve n\cdot {\hat{\ve W}})(\ve n\wedge {\hat{\ve W}})\,.
\eqnlab{torque_fluid_inertia}
\end{eqnarray}
The shape factor $F(\lambda)$ is given in Ref.~\cite{Dab15}. It is also shown in Fig.~\ref{fig:shape_factor}(a).

Combining Eqs.~(\ref{eq:eom}), (\ref{eq:Ip}), (\ref{eq:resistance}) with Eqs.~(\ref{eq:drag_correction},\ref{eq:torque_fluid_inertia}) yields the equations of motion for our model.
We use the Kolmogorov time  $\tauK$ and the Kolmogorov length $\etaK$
to de-dimensionalise the equations of motion, $\ve x'=\ve x/\etaK$, $ t'=t/\tauK$, $\ve v'=\ve v\tauK/\etaK$, $\ve\omega'=\ve\omega\tauK$.
This gives
(after dropping the primes):
\begin{subequations}
\label{eq:eom_dimless}
\begin{eqnarray}
\dot{\ve x}_{\rm p}&=\ve v_{\rm p}\,,\\
\label{eq:Drag}
\dot{\ve v}_{\rm p}&=\frac{1}{\st}\left[-\Big(\mathbbm{1}+\tfrac{3}{16}
\tfrac{a}{\etaK}W\big[3\ma A-\ma I (\hat{\ve W}\cdot \ma A \hat {\ve W})\big]\Big)\ma A \ve W +\sv\gghat\right]\,,\\
\dot{\ve n}&=\ve\omega_{\rm p}\wedge\ve n\,,\\
\label{eq:omega_p0}
\dot{{\ve\omega}}_{\rm p}&=\frac{1}{\st}\left[
\ma I_{\rm p}^{-1}\ma C(\ve\Omega-\ve\omega_{\rm p})
+\ma I_{\rm p}^{-1}\ma H\cddot\ma S
+\mathscr{A}'(\ve n\cdot\ve W)(\ve n\wedge\ve W)\right]\\
&\hspace*{1cm}+\Lambda(\ve n\cdot\ve\omega_{\rm p})(\ve\omega_{\rm p}\wedge\ve n)\,.
\nonumber
\end{eqnarray}
\end{subequations}
Eqs.~(\ref{eq:eom_dimless}) have four independent dimensionless parameters:
\begin{eqnarray}
\label{eq:dimless_parameters}
\Lambda=\frac{\lambda^2 - 1}{\lambda^2 + 1}\,,\hspace{0.5cm}
\frac{a}{\etaK}
\,,\hspace{0.5cm}
\st=\frac{\tau_{\rm p}}{\tauK}\,,\hspace{0.5cm}
\sv=\frac{g\tau_{\rm p}\tauK}{\etaK}\,.
\end{eqnarray}
Here $\Lambda$ is the shape parameter that appears in Jeffery's equation, and $\sv$ is the settling number~{\cite{Dev12}},  a dimensionless measure of the settling speed. It is proportional to the particle size squared, $a^2$, just as the Stokes number.

The  shape-dependent prefactors in \Eqnref{eom_dimless} are [in addition to those given by Eqs.~\eqnref{def_Mt} and \eqnref{eqm_C_coeffs}]
\begin{eqnarray}
&[\ma I_{\rm p}^{-1}\ma C]_{ij}=
\frac{C_\perp}{I_{\perp}} (\delta_{ij}-n_i n_j) + \frac{C_\parallel}{I_{\parallel}}n_i n_j\,,
\quad
[\ma I_{\rm p}^{-1}\ma H]_{ijk}=-\frac{C_{\perp} \Lambda}{I_{\perp}}\epsilon_{ijl} n_k n_l\,,
\end{eqnarray}
as well as
\begin{eqnarray}
\label{eq:Aprime}
&\mathscr{A}'=\frac{5}{6\pi}F(\lambda) \frac{\max(\lambda,1)^3}{\lambda^2+1}\,.
\end{eqnarray}

The Reynolds number $\re_p$ does not appear explicitly in Eqs.~(\ref{eq:eom_dimless}) because
we de-dimensionalised the equations of motion with the Kolmogorov scales $\tauK$ and $\etaK$.
If we use an estimate of the slip velocity instead (such as $W_\perp^{(0)}$),
then $\re_p$ features in the dimensionless equations of motion. The latter convention is used
in Refs.~\cite{Kha89,Dab15}, and more generally in perturbative calculations of weak inertial
effects on the motion of particles in simple flows \cite{Lovalenti93,Saf65,Saf68,Candelier2018}.
These two different choices must lead to equivalent equations of motion, but our scheme
has the advantage that it emphasises the different roles played by
$\ve f^{(1)}$ and $\ve \tau^{(1)}$ for small particles in turbulence. Eq.~(\ref{eq:Drag}) shows
that the fluid-inertia contribution to the force, $\ve f^{(1)}$, is multiplied by the dimensionless prefactor
$a/\etaK$.
This means that $\ve f^{(1)}$ makes only a small contribution for small enough particles, which we do not expect to qualitatively change the results derived below. In the following we therefore neglect this contribution (although it could be taken into account in simulations and theory).

More importantly, the fluid-inertia contribution to the torque in Eq.~(\ref{eq:omega_p0}) has no such factor. The fluid-inertia torque is of the same order
as the Jeffery torque. This implies that the fluid-inertia contribution to the
torque is potentially much more significant than the fluid-inertia contribution to the force.
At large $\sv$ in particular the particle settles rapidly, so that $\ve W$ is large.
 In this limit one therefore expects the fluid-inertia torque $\ve\tau^{(1)}$ to dominate over Jeffery's torque
 $\ve \tau^{(0)}$, so that the inertial torque cannot be neglected
 (as was done in Refs.~\cite{Siew14a,Siew14b,Gus17,Jucha2018,Naso2018}).
 It is argued in Ref.~\cite{Sha19} that  the orientation bias predicted in  Refs.~\cite{Siew14a,Siew14b,Gus17} can possibly be observed in small-$\re_\lambda$ flow, but not at high $\re_\lambda$.

 In the following we  neglect  the contribution
from $\ve f^{(1)}$. At the same time we assume that the settling speed is so   large  that
 the fluid-inertia torque $\ve\tau^{(1)}$ dominates the angular dynamics. If there was no flow, the particles would settle with their broad side first in this limit. The question is how
turbulent fluctuations modify the orientation distribution of the settling particles.

\subsection{Statistical model}
\label{sec:sm}
In our theory we use a statistical model  \cite{Gus16} to represent the turbulent fluctuations.  We model
the incompressible homogeneous and isotropic turbulent fluid-velocity field $\ve u(\ve x,t)$ as a Gaussian random function with correlation length $\ell$, correlation time $\tau$, and rms magnitude $u_0$
(here and in Section \ref{sec:KS} we write the equations in dimensional form  because we want to make explicit
how these scales are related to the Kolmogorov scales).
Following Ref.~\cite{Gus16} we express the  fluid-velocity field $\ve u(\ve x,t)$ in  three spatial dimensions (3D) as
\begin{eqnarray}
\label{eq:def_u}
\ve u &= \mathscr{N}_3 \,\ve \nabla \wedge \ve A\,.
\end{eqnarray}
The components $A_j$ of the vector field $\ve A$ are Gaussian random functions with mean zero,
$\langle A_j (\ve x,t)\rangle =0$, and with correlation functions
\begin{equation}
\label{eq:u_corr}
\langle A_i(\ve x,t)A_j(\ve x',t') \rangle=  \delta_{ij} \ell ^2 u_0^2 {\rm exp}\Big(
-\frac{|\ve x-\ve x'|^2}{2\ell^2} - \frac{|t-t'|}{\tau}\Big)\,.
\end{equation}
We choose the normalisation $\mathscr{N}_3 = 1/\sqrt{6}$ so that $u_0 = \sqrt{\langle |\ve u|^2\rangle}$.
Below we also quote results for a two-dimensional (2D) version of this model. In this case we take
\begin{eqnarray}
\label{eq:u_2d}
\ve u = \mathscr{N}_2\begin{bmatrix}
\phantom-\partial_{2} A_3\\-\partial_{1} A_3
\end{bmatrix}
\end{eqnarray}
with $\mathscr{N}_2 = 1/\sqrt{2}$, and where $\partial_j$ represents the derivative
with respect to the spatial coordinate $x_j$.  As the equation of motion for the 2D model
we use Eq.~(\ref{eq:eom_dimless}) with
 $\ve n$ and the translational dynamics constrained to the flow plane.

The statistical model has an additional dimensionless parameter, the Kubo number \cite{Gus16}
$\ku= u_0\tau/\ell$. In the limit of large $\ku$ the Gaussian random function $\ve u(\ve x,t)$ models small-scale fluid-velocity fluctuations in the  dissipative range of  homogeneous isotropic turbulence. Evaluating
Eq.~(\ref{eq:tauK}) in the statistical model gives (Section~5.1 in  Ref.~\cite{Gus16}):
\begin{eqnarray}
\label{eq:tau_def}
\frac{\tau}{\tauK}= \sqrt{d+2} \,\ku\,,
\end{eqnarray}
where $d$ is the spatial dimension.
The spatial cor\-re\-la\-ti\-on length $\ell$ satisfies $\ell^2 = \langle u_1^2 \rangle/\langle (\partial_1u_1)^2 \rangle$, which defines the Taylor length scale \cite{Fri97} in turbulence.
The length scale $\ell$  is related to the Kolmogorov length by \cite{Fri97,Cal09}
\begin{eqnarray}
\label{eq:ell}
\frac{\ell}{\etaK}
&= \mathscr{C} \sqrt{{\rm Re}_\lambda }\,,
\end{eqnarray}
where $\mathscr{C}$ is a constant of order unity. The
ratio $\ell/\etaK$ (or alternatively the Taylor-scale Reynolds number ${\rm Re}_\lambda$) constitutes a sixth dimensionless parameter of the model, in addition to the Kubo number and the four parameters listed in Eq.~(\ref{eq:dimless_parameters}). In all statistical-model simulations described
in this paper  we set $\ku=10$ and ${\ell}/\etaK=10$, and we determine the parameters $\tau$ and $\ell$
of the statistical model from Eqs.~(\ref{eq:tau_def},\ref{eq:ell}).

The statistical model is constructed to approximate the  dissipative-range fluctuations of 3D turbulence \cite{Gus16}. We note that
the predictions of the 2D and 3D statistical models are essentially similar, but the two-dimensional
model is easier to analyse, and it can be simulated more accurately.
Two-dimensional  and three-dimensional  turbulence, by contrast, exhibit significantly different fluid-velocity fluctuations.

\subsection{Kinematic-simulation model}
\label{sec:KS}
To demonstrate the robustness of our theory we also compare its predictions
to results of numerical simulations using
a different model for the turbulent flow,  namely the Kinematic-Simulation (KS) model~\cite{Fung:92}. The KS model
has been shown to reproduce qualitatively many features of turbulent transport, and it
provides a convenient way to represent a flow with a wide range of spatial scales, such as turbulence,
albeit in a simplified manner.  In short, we discretise Fourier space in geometrically spaced shells, up to a largest wavenumber. The largest and  smallest length scales of the flow are  $L$ and $\eta$, respectively.
The total number of shells is denoted by $N_k$.  We choose the characteristic wave vector in shell $n$ as:
$k_n = k_1 (L/{\eta})^{(n-1)/(N_k - 1)}$. In each cell, we pick one wave vector, ${\ve k}_n$. The flow is then simply constructed as a sum of Fourier modes:
 \begin{equation}
\ve u ( \ve x, t) = \sum_{n=1}^{N_k} {\ve a}_n \cos( {\ve k}_n \cdot{ \ve x} + \omega_n t) + {\ve b}_n \sin( {\ve k}_n \cdot \ve x + \omega_n t)  \,.
\label{eq:KS}
\end{equation}
The Fourier coefficients are chosen so that ${\ve k}_n \cdot {\bf a}_n ={\ve k}_n \cdot {\ve b}_n = 0$ (incompressibility), and with magnitude  $a_n^2 = b_n^2 = E(k_n) \Delta k_n$, where $E(k_n) = E_0 k_n^{-5/3}$ represents the Kolmogorov spectrum \cite{Fri97}. The frequency $\omega_n$ in Eq.~\eqref{eq:KS}
is taken to be $\omega_n = \tfrac{1}{2} \sqrt{ k_n^3 E(k_n) }$. Further details about the implementation of this model for $\ve u(\ve x,t)$ can be found in  Ref.~\cite{Vosskuhle:15}.

\section{Orientation distributions}
\label{sec:od}
\begin{figure}
\begin{overpic}[clip]{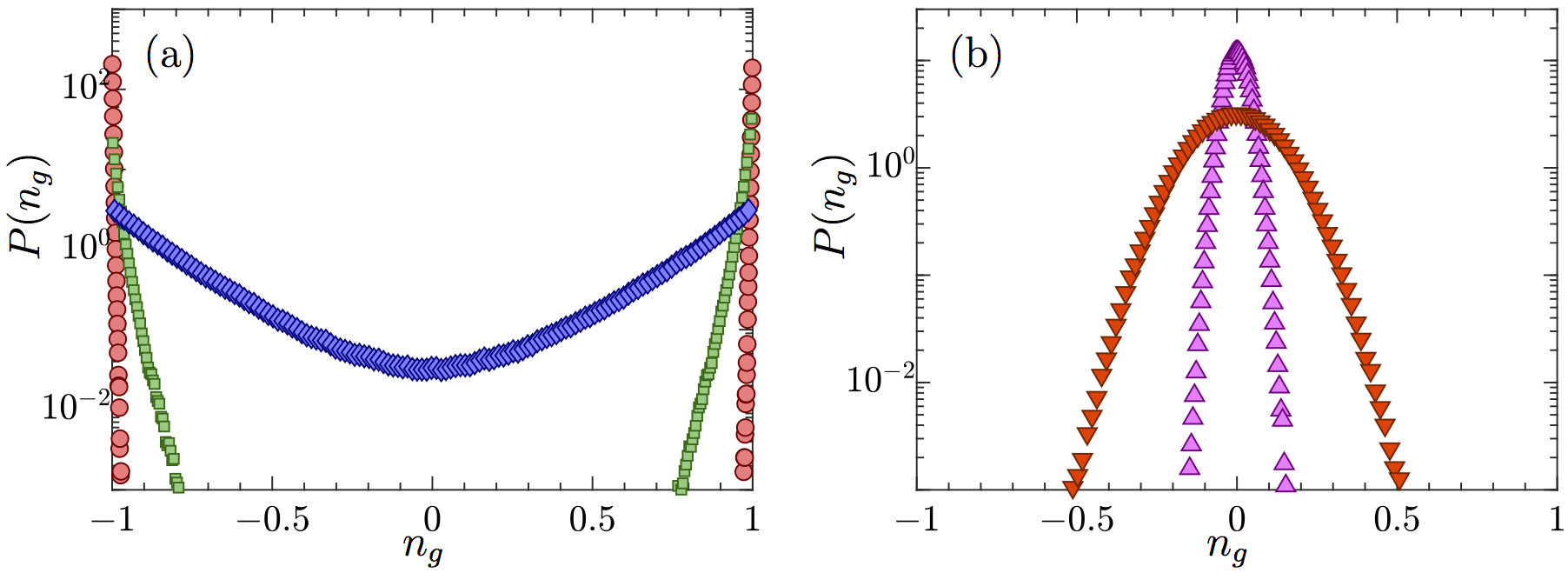}
\end{overpic}
\caption{\label{fig:distributions}  Distribution of $n_g= \ve n\cdot \gghat$ obtained by numerical simulations
of Eqs.~(\ref{eq:eom_dimless}) for the three-dimensional statistical model.
(a) Disk-like particles with aspect ratio $\lambda=0.1$, $\sv=4.5$, $\st=0.022$ (red,$\circ$), $\st=0.22$ (green,$\Box$), $\st=2.2$ (blue,$\diamond$). (a) Rod-like particle with $\lambda=5$, $\sv=45$, $\st=0.22$ (magenta,$\vartriangle$) and $\st=2.2$ (red,$\triangledown$). }
\end{figure}
Figure~\ref{fig:distributions} shows orientation distributions obtained
by numerical simulations of Eqs.~(\ref{eq:eom_dimless}) for the three-dimensional statistical model
described in Section \ref{sec:sm}. Shown are distributions of $n_g = \ve n\cdot\gghat$  for a range of different Stokes numbers.
We see that the particles settle with their broadside approximately aligned with gravity.
For rods this means that $\ve n \perp \gghat$, so that $n_g=0$, and for
disks $\ve n \parallel \gghat$, so that $n_g=1$. These are the stable orientations
for prolate and oblate particles settling in a quiescent fluid \cite{Kha89,Dab15}.

Compare the distributions in Fig.\ref{fig:distributions}  to those shown in Fig.~1 of Ref.~\cite{Gus17}.
There, by contrast, the rods
tend to settle tip first, and disks tend to settle edge first.
The reason for the difference is that
the effect of the fluid-inertia torque was neglected in Ref.~\cite{Gus17}, whereas in the present work we choose parameters
where  this torque dominates the angular dynamics.

When the Stokes number
is small we expect that the vector $\ve n$ spends most of its time close to a stable fixed point of the angular dynamics. But we expect that the turbulent velocity gradients modify the fixed point,
 so that it is no longer simply $n_g=0$ (rods) or $n_g=1$ (disks).
Since the turbulent velocity gradients change as  functions of time,
the fixed-point orientation becomes time dependent too.
 In the overdamped limit (small Stokes numbers) we expect that the particle orientation follows the fixed-point orientation quite closely.  This allows us to derive a theory for the orientation distribution in this limit, described in the following Section.

\section{Overdamped limit}
\seclab{advective}

The model (\ref{eq:eom_dimless}) is very difficult to analyse in general. Therefore, to simplify the analysis, we consider a limit of the problem where the relaxation time of  $\ve n$ is much faster than  the time scale on which the gradients change as the particle moves
through the flow. This corresponds to the overdamped limit of the problem,  $\st\to 0$  in Eqs.~(\ref{eq:eom_dimless}). It was shown by experiments and numerical simulations in Ref.~\cite{Lop17} that this limit quantitatively describes the orientation distribution of rods settling in a two-dimensional vortex flow, and in the slender-body limit this approach was also used in Refs.~\cite{Kramel,Men17}.

We also assume that $\sv$ is large enough so that
the fluid-inertia torque dominates the angular dynamics. This allows us to take into account turbulent fluctuations
perturbatively. It also means that we can approximate the instantaneous slip velocity by $\ve W^{(0)}(\ve n)$,
Eq.~(\ref{eq:Wslip}). In this limit we find:
\begin{subequations}
\begin{eqnarray}
\ve W&=\ve W^{(0)}(\ve n)\,,\\
\ve \omega_{{\rm p}}&=\ve \Omega+\Lambda (\ve n\wedge \ma S\ve n)
+{{\mathscr A}\sv^2} n_g(\ve n\wedge\gghat)\,,
\eqnlab{omega_overdamped}
\end{eqnarray}
with $n_g=\ve n\cdot\gghat$, as defined in Section \ref{sec:od}.
The overdamped equation for the dynamics of the
vector $\ve n$ corresponding to \Eqnref{omega_overdamped} reads
\begin{eqnarray}
\label{eq:odn_general}
\dot{\ve n}&=
\ma O\ve n+\Lambda[\ma S\ve n-(\ve n\cdot\ma S\ve n)\ve n]+{\mathscr A}\sv^2{n_g}(\gghat-n_g\ve n)\,.
\end{eqnarray}
\end{subequations}
To simplify the notation we introduced the parameter
\begin{eqnarray}
\label{eq:defA}
\mathscr{A} = \mathscr{A}' \frac{{I_\perp}}{A_\parallel A_\perp C_\perp}\,.
\end{eqnarray}
Fig.~\ref{fig:shape_factor}(b) shows how $\mathscr{A}$ depends on
the particle-aspect ratio $\lambda$.

\subsection{Two-dimensional dynamics in the overdamped limit}
\label{sec:2dod}
We consider the 2D model  first because it is much easier to analyse than the three-dimensional model.  We assume that the gravitational acceleration points into the $\hat{\bf e}_1$-direction, and
 define $\phi$ to be the angle ($0 \leq \phi<\pi$) between $\ve n$ and this axis, so that
 $n_g = \ve n\cdot\gghat=\cos\phi$. For prolate particles ($\lambda >1$ or equivalently $\Lambda >0$)
 the overdamped
 angular dynamics (\ref{eq:odn_general}) becomes in two spatial dimensions:
 \begin{eqnarray}
\label{eq:od4}
\tfrac{{\rm d}}{{\rm d}t}\phi&=\Omega+\Lambda[S_{12}\cos(2\phi)-S_{11}\sin(2\phi)]+\tfrac{1}{2}|{\mathscr A}|\sv^2\sin(2\phi)\,.
\end{eqnarray}
This two-dimensional overdamped equation of motion for the angular dynamics is essentially equivalent to model M2 in Ref.~\cite{Lop17}, used there for  simulations of the angular dynamics of  rods settling in a two-dimensional  vortex flow. Apart from the fact that Ref.~\cite{Lop17} considers a different flow,
it describes  small cylindrical particles with slightly different resistance tensors, and it
approximates the $\ve n$-dependence of the settling velocity.

Equation (\ref{eq:od4}) shows that the fluid-inertia torque has the same angular dependence as the $S_{11}$-component
of the strain, but in general the sign may differ. When $S_{11}>0$, the strain tends to align the rod
with $\hat {\bf e}_1$, the direction of gravity. The fluid-inertia torque acts against alignment
with this direction.  To quantify this statement,  consider the fixed points of the angular dynamics (\ref{eq:od4}).
In the limit $|\mathscr A|\sv^2\to\infty$ the inertial torque dominates the angular dynamics,
so that the fluid-velocity gradients do not matter. In this limit the fixed points are
 $\phi_1^*=0$  and $\phi_2^*=\pi/2$.
 For a prolate particle ($\lambda>1$) $\phi_1^\ast=0$ is unstable
 while $\phi_2^\ast=\pi/2$ is stable.
 This is the limit considered in Ref.~\cite{Kha89}, a  slender rod falling in a quiescent fluid:
since $\phi_2^\ast$ is stable the rod settles with its broad side first. For an oblate particle
the stabilities are reversed \cite{Dab15}.
\begin{figure}
\begin{overpic}[clip]{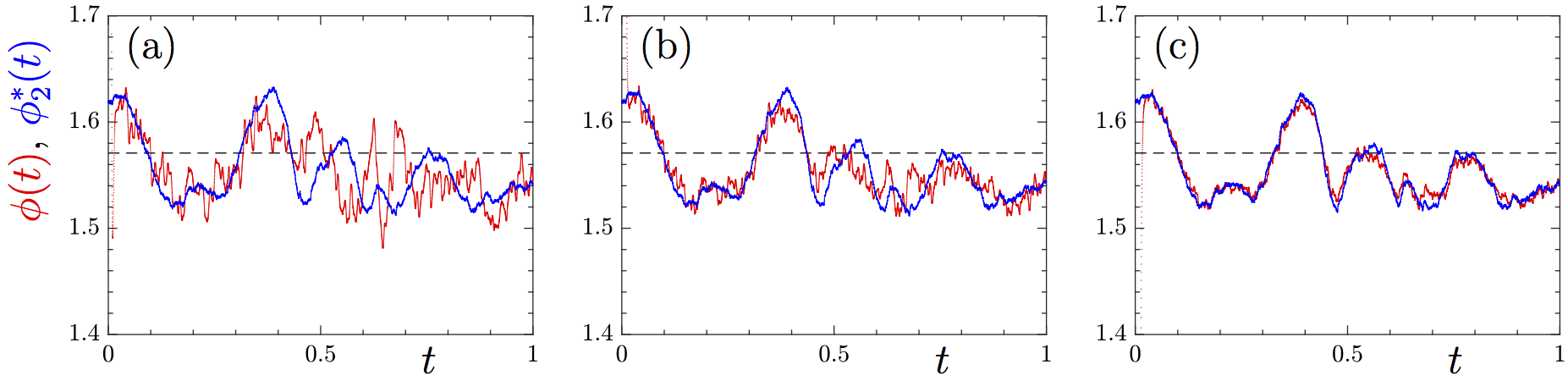}
\end{overpic}
\caption{\figlab{angular_dynamics} Angular dynamics of a settling particle in two spatial dimensions.
Shown is the angle $\phi(t)$ obtained by simulation of Eqs.~(\ref{eq:eom_dimless}) (red), and the analytically exact result for the stable fixed
point $\phi_2^\ast(t)$ (blue). (a) ${\st} = 0.1$, (b) ${\st}=0.05$, (c) ${\st}=0.02$. Other parameters: ${\rm Sv} = 25$, $\lambda =5$.
The three simulations were performed with the same initial conditions and for the same realisation of the function
$\ve u(\ve x,t)$ in the 2D statistical model.}
\end{figure}

What is the effect of the turbulent flow? In general this question is difficult to answer.
But if the angle $\phi$ relaxes much more quickly than the fluid-velocity gradients change along the particle path, then  the problem becomes tractable.
Assuming that the gradients are constant, we can find exact
expressions for the  two fixed points of \Eqnref{od4},
for arbitrary aspect ratios and fluid-velocity gradients.
We take $\lambda>1$ and expand the stable fixed point around $\pi/2$ assuming that  $|\mathscr A|\sv^2$ is large:
\begin{eqnarray}
\phi_2^* &=\frac{\pi}{2}-{B_{12}\frac{1}{|{\mathscr A}|\sv^2}-2B_{11}B_{12}\frac{1}{({\mathscr A}\sv^2)^2}+\dots}
\label{eq:phi_expand}
\end{eqnarray}
Here $B_{ij}$ are the elements of the matrix $\ma B = \ma O + \Lambda \ma S$.
Eq.~(\ref{eq:phi_expand}) shows how the fixed-point orientation changes as the turbulent velocity gradients evolve.
We expect that the orientation of a settling rod follows these fixed-point orientations closely in the overdamped limit, provided that its angular relaxation time is smaller than the time scale on which the flow (and thus $\phi_2^\ast$) changes. We now analyse the angular dynamics of the settling particles in this \lq{}persistent limit\rq{}     \cite{Mei19}.

\Figref{angular_dynamics} shows examples of how the fixed point $\phi_2^\ast(t)$ of the angular dynamics
fluctuates
as the particle settles through the turbulent flow and encounters different fluid-velocity gradients. The data are obtained by numerical simulation of the 2D model described in  Section \ref{sec:formulation}, for small Stokes numbers.
Also shown is the instantaneous angle $\phi(t)$ obtained in these simulations. We see that the orientation
dynamics follows the fixed point $\phi_2^*$ quite closely when $\st$ is small.

In the overdamped limit the relaxation time $\tau_\phi$  of the angular dynamics (in units of $\tauK$)  is given by the inverse of the  stability exponent $\sigma$ of the fixed point $\phi_2^*$. From (\ref{eq:od4}) we find to first order in $(|\mathscr A|\sv^2)^{-1}$ that $\sigma\sim {-}|{\mathscr A}|\sv^2$. This gives
\begin{eqnarray}
\tau_\phi\sim{|\sigma^{-1}|}&=\frac{1}{|{\mathscr A}|\sv^2}\,.
\eqnlab{tau_overdamped}
\end{eqnarray}
When $\sv$    is large, the fluid-velocity gradients seen by the settling particle
change at the settling time scale $\tauL$, 
 the time it takes a particle settling with an angle $\phi=\pi/2$ at a settling velocity given by \Eqnref{Wslip} to fall one correlation length $\ell$
\begin{eqnarray}
\tauL&=\frac{1}{\tauK}\frac{\ell A_\perp}{\tau_{\rm p}{g} }=\frac{\ell}{\etaK}\frac{A_\perp}{\sv}\,.
\eqnlab{tau_settling}
\end{eqnarray}
We therefore conclude that the persistent limit requires:
\begin{eqnarray}
\label{eq:ratio}
\frac{\tau_\phi}{\tauL }&= {\frac{1}{{A_\perp}|{\mathscr A}|\sv}} \frac{\etaK }{\ell }\ll 1\,.
\end{eqnarray}
This indicates that the persistent approximation
works in the overdamped limit when $|\mathscr{A}|\sv$ is large enough.
In the opposite limit, for small values of $\sv$, the settling time scale $\tauL$ is larger than 
the Lagrangian time scale, so that the fluid-velocity gradients change at the Lagrangian time scale, of order unity
in units of $\tauK$.
Hence we must demand  $\tau_\phi \ll 1$ to ensure that the persistent approximation works. This corresponds to the condition
\begin{eqnarray}
\label{eq:persistent_condition_small_Sv}
    |\mathscr{A}|\sv^2\gg 1\,.
\end{eqnarray}
%% It fails when  $\sv$ becomes so small that (\ref{eq:persistent_condition_small_Sv}) is no longer satisfied.
In the persistent limit, the overdamped angular dynamics (\ref{eq:od4})
responds so  rapidly
 that the orientation of the particle  follows the instantaneous fixed point of the dynamical system
 (\ref{eq:od4}) quite closely.
In this case the orientation distribution of the settling particle is determined by the distribution
of $\phi_2^\ast$, and thus by the distribution
of fluid-velocity gradients encountered by the particle,  through Eq.~(\ref{eq:phi_expand}). This distribution may differ from the
distribution of fluid-velocity gradients at a fixed spatial position (preferential sampling \cite{Gus16}).
But in the overdamped limit preferential sampling of the fluid-velocity gradients is expected to be weak. We have
checked that it is negligible for data  shown in this paper.
\begin{figure}[t]
\mbox{}\\[4mm]
\begin{overpic}[clip]{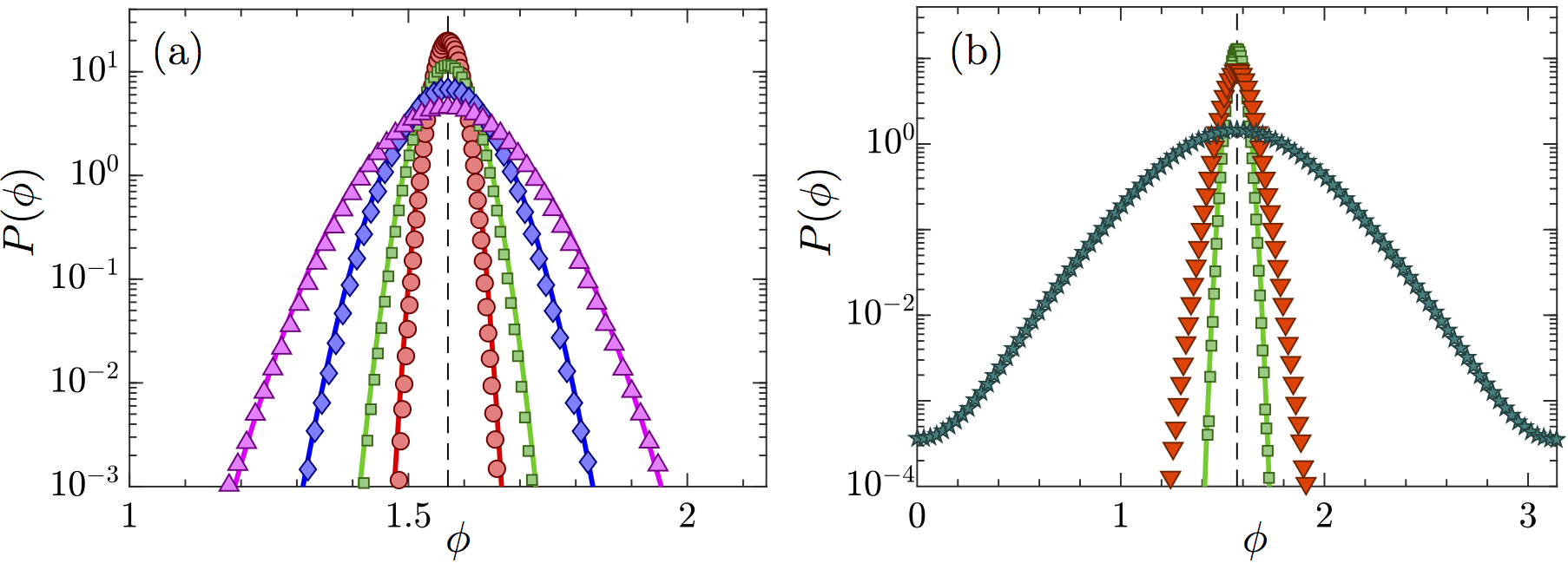}
\end{overpic}\caption{\label{fig:od_2D}
Orientation distributions for the two-dimensional statistical model.
(a)
Distribution of angle $\phi={\rm acos}(n_g)$ obtained from numerical simulation of the dynamics \eqnref{eom_dimless} (markers) and the limiting theory for small Stokes numbers, Eq.~\eqnref{pphi4} (solid lines). Parameters:
{ ${\sv}=22$, ${\st}=0.022$}, and $\lambda=3$ (red,$\circ$), $\lambda=5$ (green,$\Box$), $\lambda=7.5$ (blue,$\Diamond$), $\lambda=10$ (magenta,$\vartriangle$).
(b) Same, but for different Stokes numbers. Parameters: $\lambda=5$, and ${\st}=0.022$ (green,$\Box$), ${\st}=0.22$ (red,$\triangledown$), ${\st}=22$ (dark green,$\star$). }
\end{figure}

If we consider only the leading correction in Eq.~(\ref{eq:phi_expand}), then the orientation distribution
is determined by the distribution $P_B(B_{12})$ of $B_{12}$:
\begin{eqnarray}
\label{eq:pphi3}
P(\phi)&\!=\!\int_{-\infty}^{\infty}\!\!\!\!{\rm d}B_{12}\,P_B(B_{12})\,\,\delta\Big(\phi\!-\!\frac{\pi}{2}+\frac{B_{12}}{|{\mathscr A}|\sv^2}\Big)=P_B\big[(\tfrac{\pi}{2}\!-\!\phi)|\mathscr{A}|\sv^2\big]\,.
\end{eqnarray}
In the two-dimensional statistical model the distribution $P_B(B_{12})$ is Gaussian
with variance $\sigma_{{B}}^2=\tfrac{1}{{8}}(2+\Lambda^2)$.
This means that  the distribution of $\phi$ is Gaussian too:
\begin{eqnarray}
\label{eq:pphi4}
P(\phi)
=\frac{{\rm e}^{-\frac{(\phi-\pi/2)^2}{2\sigma_\phi^2}}}{\sqrt{2\pi\sigma_\phi^2}}\,,
\end{eqnarray}
with variance
\begin{eqnarray}
\sigma_{\phi}^2=
\frac{1}{{8}}\frac{2+\Lambda^2}{(|{\mathscr A}|\sv^2)^2}\,.
\eqnlab{PphiTheory}
\end{eqnarray}
Eq.~(\ref{eq:pphi3}) shows that the distribution of $\phi$ simply reflects that of the fluid-velocity gradients,
in the overdamped and persistent limit.
The corresponding distribution of $n_g= \ve n\cdot\hat{\ve g}$ is:
\begin{eqnarray}
\label{eq:P_ng_2d}
P(n_g)&=\frac{1}{\sin\phi}P(\phi)
=\frac{\exp\left[-({\rm acos}(n_g)-\pi/2)^2/(2\sigma_\phi^2)\right]}{\sqrt{2\pi\sigma_\phi^2}\sqrt{1-n_g^2}}\,.
\end{eqnarray}
Fig.~\ref{fig:od_2D} shows that Eqs.~\eqnref{pphi4}  and \eqnref{PphiTheory} agree well with results of simulations of the overdamped dynamics in two spatial dimensions, provided
that $\st$ is small enough [panel (a)]. When the Stokes number becomes larger [panel (b)], the distribution
is much wider than predicted by the overdamped theory.

\subsection{Three-dimensional dynamics in the overdamped limit}
\label{sec:threedimoverdamped}
In this Section we show how to obtain the distribution of $n_g=n\cdot \hat{\ve g}$ for the three-dimensional
statistical model, in the same overdamped and persistent limit considered above. The calculation
is analogous to the one described in Section \ref{sec:2dod}.
Let $\ve p=\ve n-n_g\hat{\ve g}$. Using $p^2=1-n_g^2$
we express
the equation of motion {(\ref{eq:odn_general})} of  $n_g$ as
\begin{eqnarray}
\nonumber
\dot n_g&=\hat{\ve g}\cdot \dot{\ve n}
=\hat{\ve g}\cdot\ma O\ve n+\Lambda[\hat{\ve g}\cdot \ma S\ve n-(\ve n\cdot\ma S\ve n)n_g]+{{{\mathscr A}\sv^2}n_g}(1-n_g^2)\\
&\!=\!O_{gp}\!+\!\Lambda[(1\!-\!2n_g^2)S_{gp}+n_g(1\!-\!n_g^2)S_{gg}\!-\!n_gS_{pp}]
+{{{\mathscr A}\sv^2}n_g}(1-n_g^2)\,.
\label{eq:expand_9}
\end{eqnarray}
Here the subscripts $g$ and $p$ denote contractions with $\hat{\ve g}$ and $\ve p$.
In the limit of $|\mathscr A|\sv^2\to\infty$, $n_g^\ast=0$ is the stable fixed point for
prolate particle ($\lambda >1$). To determine how the fixed point changes due to fluid-velocity
fluctuations we seek an expansion in $(|{\mathscr A}|\sv^2)^{-1}$ as in Section \ref{sec:2dod}, of the form
$n_g^*\propto1/(|{\mathscr A}|\sv^2)+\ldots$.
We obtain to leading order:
\begin{eqnarray}
\label{eq:fpng}
n_g^*&=\frac{\hat{\ve g}\cdot \ma B\ve p}{|{\mathscr A}|\sv^2}\,.
\end{eqnarray}
Assuming that the orientation of $\ve p$ is  uncorrelated from the
fluid-velocity gradients, we obtain for the variance
\begin{figure}
\mbox{}\\[4mm]
\begin{overpic}[clip]{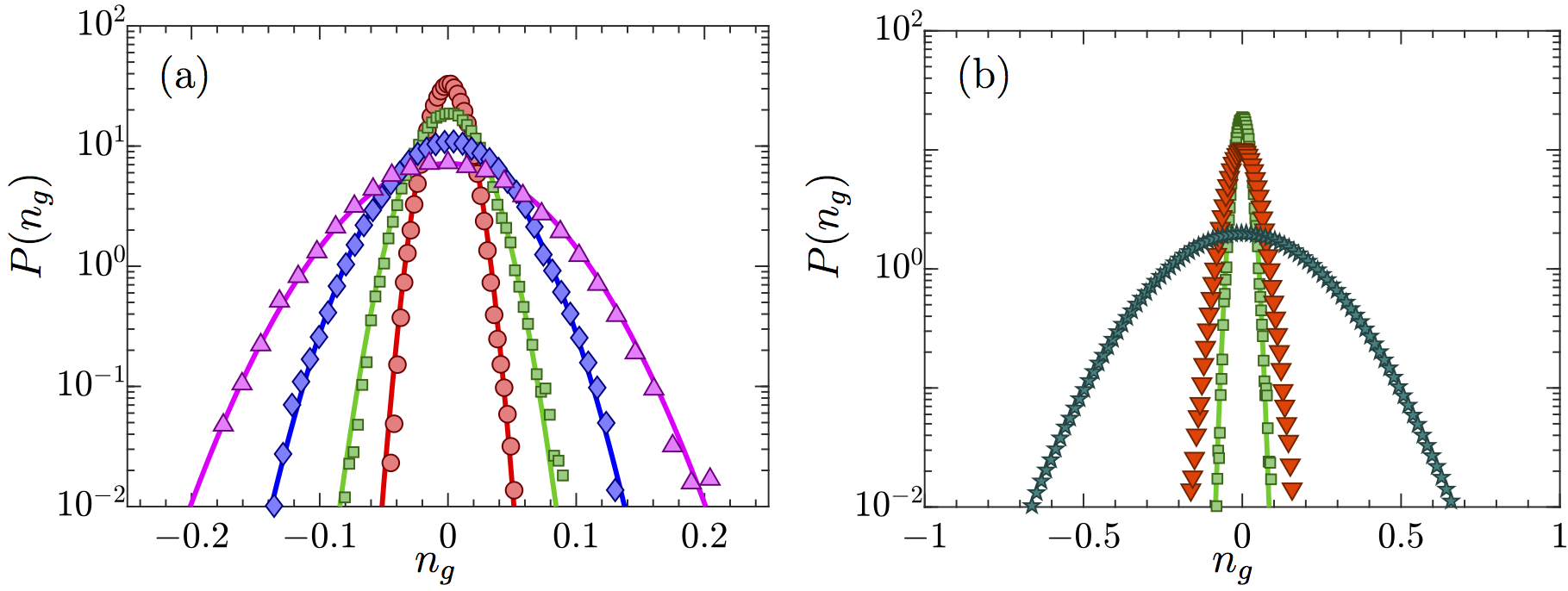}
\end{overpic}
\caption{\label{fig:3Ddist_rods} Orientation distribution for the three-dimensional
statistical model. Same conventions and parameters as in \Figref{od_2D}. (a) $P(n_g)$ in the overdamped limit.
(b) Same, but for different Stokes numbers.
}
\end{figure}
 of the distribution of $n_g$:
\begin{eqnarray}
\label{eq:var_general}
\sigma_{n_g}^2 = \frac{\langle B_{12}^2 \rangle\langle |\ve p|^2 \rangle }{(|\mathscr{A}|\sv^2)^2}
\approx  \frac{ \sigma_B^2 }{(|\mathscr{A}|\sv^2)^2}\,,
\end{eqnarray}
where
$\sigma_B^2$ is the variance of the distribution
of $B_{12}$ (the gravitational acceleration points in the $\hat{\bf e}_1$-direction). We also  used that  $p^2=1-n_g^2\approx 1$. This is a good approximation
because in the limit we consider $n_g$ is small  for prolate particles.
Assuming that $\ve p$ and the fluid-velocity gradients are uncorrelated implies that the distribution of $n_g$ is Gaussian in the statistical model:
\begin{eqnarray}
\label{eq:dist3d}
P(n_g)&=\frac{1}{\sqrt{2\pi}\sigma_{n_g}}\exp\Big(-\frac{n_g^2}{2\sigma_{n_g}^2}\Big)\,,
\end{eqnarray}
and the variance evaluates to
\begin{eqnarray}
\label{eq:var3d}
\sigma_{n_g}^2&={\frac{1}{(|{\mathscr A}|\sv^2)^2}\frac{5 + 3\Lambda^2}{60}}\,.
\end{eqnarray}
Figure \ref{fig:3Ddist_rods} shows results for the distribution of $n_g$ from simulations of
the three-dimensional statistical model. Panel (a) shows results for
small Stokes numbers, the parameters are the same as in Fig.~\ref{fig:od_2D}(a).  Also shown are the results of the
theory, Eqs.~(\ref{eq:dist3d}) and (\ref{eq:var3d}). In this case $\st$ is small enough and $\sv$ large enough
so that the theory works very well.
Panel (b) shows the orientation distribution for different Stokes numbers, to demonstrate
how the theory fails when the Stokes number becomes larger.
The behaviour is similar to that described in Section \ref{sec:2dod}: the distribution widens as $\st$ increases.

\begin{figure}
\begin{overpic}[clip]{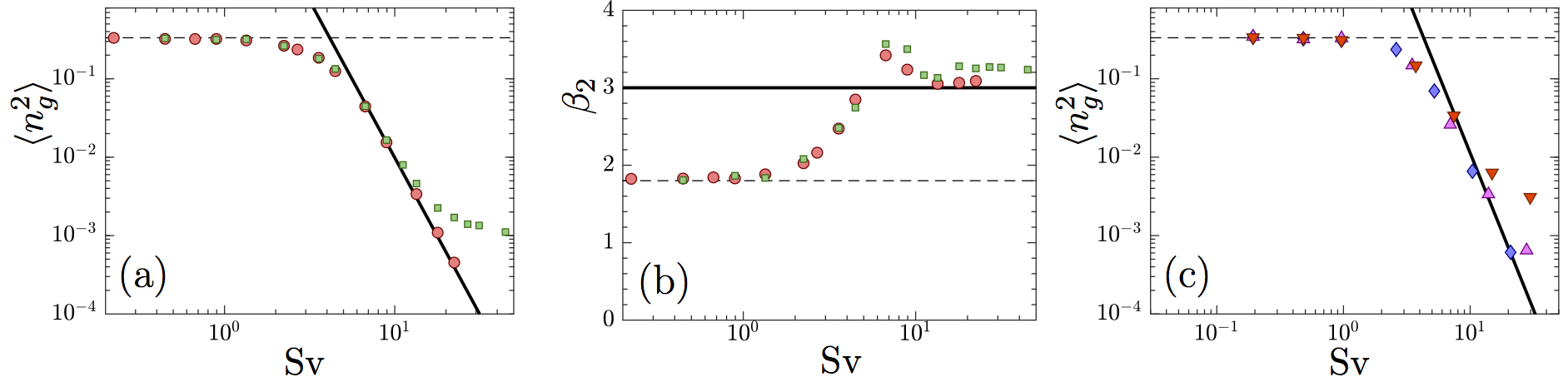}
\end{overpic}
\caption{\label{fig:variance}  Width of the orientation distribution. (a) Variance of $n_g$ from simulations
of the three-dimensional model, as a function of $\sv$, for two values of the Stokes number:
$\st=0.022$ (red, $\circ$) and $\st=0.22$ (green, $\Box$). Also shown is
the theory for large $\sv$, Eq.~(\ref{eq:var3d}), solid line,
and the result for a uniform distribution, $\langle n_g^2\rangle= \tfrac{1}{3}$ (dashed line).
(b) Kurtosis  $\beta_2=\langle n_g^4\rangle/\langle n_g^2\rangle^2$.
Same parameters as in panel (a). The
overdamped theory (Section \ref{sec:threedimoverdamped}) gives a Gaussian
distribution with kurtosis equal to $\beta_2=3$ (solid  line).
For a uniform distribution, $\beta_2= \tfrac{9}{5}$ (dashed line).
{(c) Results for $\sigma^2_{n_g}$ from KS  for $\st=0.025$ (blue,$\diamond$), $0.1$ (magenta,$\vartriangle$), and $0.4$ (red, $\triangledown$). Also shown is the theory, Eq.~(\ref{eq:var_general}), solid line, as well as the uniform limit (dashed line).}}
\end{figure}

  Eq.~(\ref{eq:var_general}) says that the variance of the distribution of $n_g$ is inversely proportional to the fourth power of $\sv$, $\sigma_{n_g}^2 \propto \sv^{-4}$,
 for large values of the settling number provided that the Stokes number is small enough. In Fig.~\ref{fig:variance}(a) this  prediction
is compared with results of simulations of the three-dimensional statistical model.
Shown is the variance of $n_g$ as a function of $\sv$, for two Stokes numbers.
When the Stokes number is small we see that the prediction (\ref{eq:var3d})
works well for large $\sv$, as expected.  Fig.~\ref{fig:variance}(b) shows the kurtosis $\beta_2=\langle n_g^4\rangle/\langle n_g^2\rangle^2$,
measuring the flatness of the distribution $P(n_g)$.
As predicted by the theory, the kurtosis approaches the Gaussian limit  ($\beta_2=3$)
for large settling numbers, at small enough Stokes numbers.

When $\sv\to 0$ the variance tends to $\tfrac{1}{3}$ and $\beta_2\to \tfrac{9}{5}$, indicating that
the persistent approximation fails because Eq.~(\ref{eq:persistent_condition_small_Sv}) is no longer satisfied. In this limit
the distribution of $n_g$ becomes uniform and independent of the Stokes number, because
the angular dynamics is isotropic  when gravitational settling is weak.
Fig.~\ref{fig:variance}(c) shows results for the variance from numerical
simulations using the KS model (Section \ref{sec:KS}), for three different values of the Stokes number.
The results are very similar to those obtained using the statistical model  [Fig.~\ref{fig:variance}(a)].
There is good agreement with the overdamped theory,  Eq.~(\ref{eq:var_general}), at large $\sv$ for
small enough $\st$. We determined
$\sigma_B^2$ from the KS simulations, so there are no fitting parameters in Fig.~\ref{fig:variance}(c). The good agreement
shows that the overdamped theory is robust, insensitive to the details of the spectrum of the velocity
fluctuations.
Fig.~\ref{fig:variance} also shows numerical data for
 larger values of $\st$. For small $\sv$ this makes little difference, the distribution
 is uniform. For larger $\sv$ the numerical results first follow Eq.~(\ref{eq:var_general}) or (\ref{eq:var3d}). But as $\sv$ increases further,
 the overdamped theory starts to fail, the earlier the larger the Stokes number. This indicates that particle inertia begins to become important.

\section{Effect of particle inertia}
\label{sec:underdamped}
We saw in the previous Section that the overdamped theory breaks down
at large $\sv$. 
%% Even for a small Stokes number this happens for large enough $\sv$, as Fig.~\ref{fig:variance} shows.
To understand when and why the overdamped theory fails one must check the full inertial dynamics.
We analyse the 2D statistical  model first.

\subsection{Two-dimensional model}
Consider the angular dynamics
in the absence of  flow, to estimate the time scales that are important
for the angular dynamics.
When $\ve u=0$, the dynamics of the phase-space coordinate
${\ve z}\equiv(v_{{\rm p}x},v_{{\rm p}y},\phi,\omega)$ has the stable fixed point  ${\ve z}^*=({\sv/A_\perp},0,\pi/2,0)$,
gravity in the direction of $\hat{\bf e}_1$.
The stability  matrix follows from Eq.~(\ref{eq:eom_dimless}):
\begin{eqnarray}
\ma J\equiv \frac{\partial\dot{{\ve z}}}{\partial{\ve z}}
=\frac{1}{\st}\left[\begin{array}{cccc}
-A_\perp & 0 & 0 & 0\cr
0 & -A_\parallel & \frac{A_\parallel-A_\perp}{A_\perp}\sv & 0\cr
0 & 0 & 0 & \st\cr
0 &{-} \frac{\mathscr{A}'}{A_\perp}\sv & {+}\frac{\mathscr{A}'}{A_\perp^2}\sv^2 & -\frac{C_\perp}{I_{\perp}}
\end{array}\right]\,,
\eqnlab{stability_inertial}
\end{eqnarray}
where $\mathscr{A}'$ was defined in Eq.~(\ref{eq:Aprime}).
The relaxation time following from Eq.~(\ref{eq:stability_inertial}) is given by
$\tau_\phi=\max(-1/\Re\sigma_i)$, the maximal stability time of $\ma J$.
Here $\Re\sigma_i$ denotes the real part of the $i$-th eigenvalue of $\ma J$.
One eigenvalue of this matrix is $\sigma= -A_\perp/\st$. We have computed
the other eigenvalues  numerically and analytically in limiting cases. We find that
 the time scale $\tau_\phi$ interpolates between \Eqnref{tau_overdamped} for small $\st$ and ${\sim}\st/A_\perp$ for large $\st$, for a fixed value of $\sv$. If we fix $\st$, by contrast, then
we find that the time scale $\tau_\phi$  interpolates between Eq.~(\ref{eq:tau_overdamped}) for small $\sv$ and ${\sim}\st/A_\perp$ for large $\sv$.

\begin{figure}
\begin{overpic}[clip]{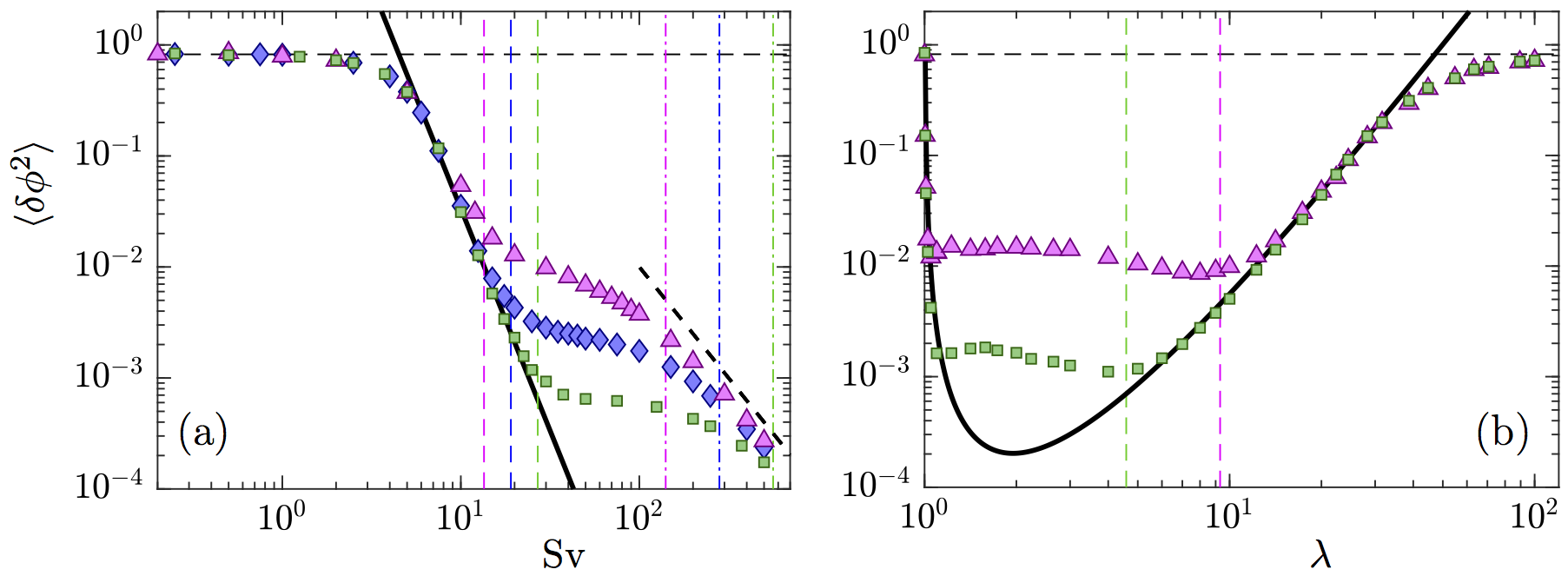}
\end{overpic}
\caption{\label{fig:variance_2D}  Variance $\langle \delta\phi^2\rangle$
 for the two-dimensional statistical model. (a) Results of numerical simulations as a function of $\sv$ for $\lambda=5$, $\st=0.1$ (green, $\Box$), $\st=0.2$ (blue, $\diamond$),
$\st=0.4$, (magenta, $\vartriangle$). Also shown: theory from Section \ref{sec:2dod}, Eqs.~\eqnref{pphi4}  and \eqnref{PphiTheory}, thick solid black line;
condition $|\mathscr{A}|\sv^2 = A_\perp/\st$ for the overdamped theory to fail [Eq.~(\ref{eq:inertial_condition})], vertical dashed lines; condition (\ref{eq:ratio2}) for
the white-noise limit, vertical dash-dotted lines; large-$\sv$ scaling (\ref{eq:ultimate}), thick black dashed line; uniform distribution at
small $\sv$, horizontal black dashed line. (b) Results as a function of the particle aspect ratio $\lambda$ for
$\sv = 25$, $\st=0.1$ (green, $\Box$),  and $\st=0.4$ (magenta, $\vartriangle$).
%%  The volume of the particle is kept fixed as $\lambda$ is varied.
}
\end{figure}
We expect that the overdamped approximation fails when the inertial estimate  for the relaxation time
of the angular dynamics, $\tau_\phi \sim \st/A_\perp$, becomes
larger than the overdamped estimate \Eqnref{tau_overdamped}. This means that the overdamped approximation
requires
\begin{align}
\label{eq:inertial_condition}
|\mathscr{A}|\sv^2 \ll A_\perp/\st\,.
\end{align}
Conversely, when Eq.~(\ref{eq:inertial_condition}) is not satisfied then particle inertia matters, so that the over\-dam\-ped
approximation must fail [Fig.~\ref{fig:variance}(a)].
For a quantitative comparison, Fig.~\ref{fig:variance_2D}(a) shows numerical results for the variance of the orientation distribution
obtained from simulations of the two-dimensional model. 
%% The condition (\ref{eq:inertial_condition}) is represented by vertical dashed lines.
We see that the overdamped approximation breaks down for values of $\sv$ larger than $\sim \sqrt{A_\perp/(|\mathscr{A}|\st)}$, as predicted by Eq.~(\ref{eq:inertial_condition}). We observe that the variance decreases more slowly
as $\sv$ increases further.

Fig.~\ref{fig:variance_2D}(a) also reveals that there is yet another,  asymptotic regime at very large values
of $\sv$ -- so large that it is difficult to achieve small $\re_p$ at the same time (Section~\ref{sec:conclusions}). It is nevertheless
of interest to analyse this regime, because it reveals  the  ingredients that a theory describing  effects of particle inertia must contain.
Fig.~\ref{fig:variance_2D}(a) suggests that
\begin{eqnarray}
 \label{eq:ultimate}
 \langle n_g^2\rangle& \sim\frac{c_1}{\sv^2}
 \end{eqnarray}
 for very large values of $\sv$. Our simulations indicate that the prefactor $c_1$ depends upon $\ell/\etaK$, $\st$, and upon $\lambda$ (not shown).
We  surmise that
this regime describes particles settling
 so rapidly that the settling time scale $\tauL$
is the smallest time scale in the system.
This cannot hold unless $\tau_\phi \sim \st/A_\perp$ is much larger than  $\tauL$,
and this crossover occurs at
\begin{eqnarray}
\label{eq:ratio2}
& \frac{\sv\,\st}{A_\perp^2} \frac{\etaK }{\ell }\sim1\,.
\end{eqnarray}
We expect Eq.~(\ref{eq:ultimate}) to be accurate for values of $\sv$ much larger than
those given by  Eq.~(\ref{eq:ratio2}).
This condition is also shown in Fig.~\ref{fig:variance_2D}, and we see that the large-$\sv$ regime starts
at values of $\sv$ approximately satisfying (\ref{eq:ratio2}). Since condition (\ref{eq:inertial_condition}) is violated in this regime,
particle inertia must be taken into account.
A difficulty is that  particle inertia changes the translational  as well as the angular dynamics.
Thus it is no longer guaranteed that $\ve W = \ve W^{(0)}(\ve n)$ (assumed  in the overdamped theory of Section \ref{sec:advective}).
This means that particle
inertia is expected to modify the angular dynamics in at least two ways. Firstly, it introduces the time derivative $\tfrac{{\rm d}^2}{{\rm d}t^2}\delta\phi$ into the angular dynamics. Secondly, the fluctuations of the torque change  because $\ve W \neq \ve W^{(0)}(\ve n)$ when particle inertia matters. This is discussed in Section \ref{sec:sa}.

Fig.~\ref{fig:variance_2D}(b) shows how the variance of $\delta\phi$ depends on particle shape, for fixed $\sv$ and $\st$. 
There are four regimes. First, in the limit $\lambda\to\infty$ the distribution is uniform and independent
of the Stokes number. In this regime the dynamics is overdamped [condition (\ref{eq:inertial_condition})],
but the persistent approximation fails because Eq.~(\ref{eq:persistent_condition_small_Sv}) is not satisfied.
Second, for intermediate aspect ratios,  both conditions are satisfied, so that the theory [Eqs.~\eqnref{pphi4}  and \eqnref{PphiTheory}] is  accurate. 
Third, at $\lambda$ becomes smaller, the overdamped approximation breaks down.
In this regime particle inertia must be taken into account. Fourth, as $\lambda\to 1$ the orientation distribution must become uniform.  This cross-over happens very rapidly: for spheres ($\lambda=1$) the orientation distribution is uniform, but already for $\lambda \sim 1.05$ there is strong alignment.

 \subsection{Klett's small-angle expansion}
 \label{sec:sa}
 Klett \cite{Kle95}  proposed a theory for the orientation variance of nearly spherical particles settling
 in turbulence, including particle inertia in the angular dynamics. He uses that
the orientation variance  is very small for large values of $\sv$. This suggests to  expand the equations of motion
in small deviations of  the angle $\phi={\rm acos}(\ve n\cdot\hat{\ve g})$ from its equilibrium value:
$\phi = \phi^\ast+ \delta\phi$ where $\phi^\ast=\tfrac{\pi}{2}$ for prolate particles.
%% Klett \cite{Kle95}  used such an expansion to analyse  the angular dynamics of nearly spherical particles settling in turbulence.  He concluded  that the orientation variance scales as $\sv^{-2}$ at large $\sv$.
%%  This is the same scaling as shown in Fig.~\ref{fig:variance_2D} for large $\sv$, and in  Eq.~(\ref{eq:ultimate}).
%% But we now show that Klett's theory cannot describe the large-$\sv$ regime discussed in the previous Section.
Klett assumes that  $\ve W=\ve W^{(0)}(\ve n)$ [\Eqnref{Wslip}] and expands the angular dynamics
for nearly spherical particles in $\delta\phi$. 

We can derive an equation of motion
consistent with his by expanding Eqs.~(\ref{eq:eom_dimless}) to leading order in $\delta\phi$,
 assuming that $\ve W=\ve W^{(0)}(\ve n)$,
 and retaining
only the leading terms in $(|\mathscr{A}|\sv^2)^{-1}$.
In this way we obtain for a prolate particle of arbitrary aspect ratio in three spatial dimensions:
\begin{eqnarray}
\eqnlab{alpha_approx}
\frac{{\rm d}^2}{{\rm d}t^2}\delta\phi&+\frac{C_\perp}{I_\perp\st}\frac{{\rm d}}{{\rm d}t}\delta\phi+\frac{C_\perp}{I_\perp\st}|\mathscr{A}| \sv^2}\delta\phi={-\frac{C_\perp}{I_\perp\st}\,\hat{\ve g}\cdot\ma B \ve p\,.
\end{eqnarray}
When we expand the geometrical coefficients
in  Eq.~\eqnref{alpha_approx} for small $\Lambda$
we find that the prefactors of the terms on the l.h.s.
of this equation
are almost identical, in this limit, to those in Eq.~(17) of Ref.~\cite{Kle95}.
Slight discrepancies arise in the $\delta\phi$-term because we use the expression for the inertial torque from Ref.~\cite{Dab15}, while
Klett uses the form obtained by Cox \cite{Cox65} (the relative error of the prefactors is of the order of $10^{- 3}$ \cite{Dab15}).
At any rate, Eq.~\eqnref{alpha_approx} is simply a damped driven harmonic oscillator, with implicit  solution
\begin{align}
\label{eq:sol}
\delta\phi(t)=\frac{C_\perp}{\Omega_0 I_\perp\st}\int_0^{t}{\rm d}t_1{\,\rm e}^{C_\perp({t_1-t})/(2I_\perp\st)}\sin[\Omega_0({t_1-t})]\,\,\hat{\ve g}\cdot \ma B(t_1)\ve p\,.
\end{align}
Here $\Omega_0=[C_\perp/(2I_\perp\St) ]\sqrt{4|\mathscr A|\sv^2I_\perp\st/C_\perp- 1}$.
Note that we discarded terms related to the initial angle, because they cannot be important
for the steady-state variance of $\delta\phi$ in the limit of large $\sv$.
Squaring Eq.~(\ref{eq:sol}) and averaging over realisations of the turbulent fluctuations in the statistical model we obtain
for large $\sv$
\begin{align}
\label{eq:sv4}
\langle \delta\phi^2\rangle {\sim \frac{c_0}{ \sv^{4}}}\,,
\end{align}
where $c_0$ is a function of $\ell/\etaK$, $\st$, and of the aspect ratio $\lambda$.
We  neglected a $\sv^{-3}$ contribution to $\langle \delta\phi^2\rangle$ because it is exponentially suppressed.
Eq.~(\ref{eq:sv4}) fails to describe the large-$\sv$ behaviour (\ref{eq:ultimate}), shown as  the thick black dashed line in Fig.~\ref{fig:variance_2D}. This means that Eq.~(\ref{eq:alpha_approx}) cannot be used to
estimate the large-$\sv$ width of the orientation distribution, or to  compute deviations from the overdamped theory. 

Which approximation causes Eq.~(\ref{eq:alpha_approx}) to fail?
Since the variance is  small for large $\sv$, $\delta\phi$ remains
small at all times. Therefore we see  no reason to doubt that the small-angle expansion
is valid. This leads us to conclude that the assumption $\ve W = \ve W^{(0)}(\ve n)$
breaks down, in agreement with our conclusions in the previous Section.
To check this, we artificially imposed the constraint $\ve W = \ve W^{(0)}(\ve n)$ in simulations of the two-dimensional
statistical model. The resulting  large-$\sv$ variance follows Eq.~(\ref{eq:sv4}), and thus fails to give the correct scaling, Eq.~(\ref{eq:ultimate}). This demonstrates that it is important to allow $\ve W$ to deviate
from $\ve W^{(0)}(\ve n)$ when particle inertia matters.

Klett's theory is difficult to justify from first principles because it assumes  that $\ve W = \ve W^{(0)}(\ve n)$.
 However, he obtains that $\langle \delta\phi^2\rangle \propto \sv^{-2}$, assuming that the fluid-velocity gradients
on the r.h.s. of Eq.~(\ref{eq:alpha_approx}) are just white noise in time.
In view of Eq.~(\ref{eq:ratio2}) it is possible that a first-principles theory may yield just that.
But fluctuations of $\ve W - \ve W^{(0)}(\ve n_t)$ yield additional time-dependent terms in the angular equation of motion
that are expected to change the properties of the noise driving the angular dynamics, resulting in a different
prediction for the orientation variance.
More importantly, Fig.~\ref{fig:variance_2D}  demonstrates that $\langle \delta\phi^2\rangle \propto \sv^{-2}$ applies only in the unphysical limit of very large
$\sv$, and that particle inertia causes a complex parameter dependence of the orientation variance at
smaller values of $\sv$, with a number of different regimes to consider.

\section{Conclusions}
\label{sec:conclusions}
Convective fluid inertia
affects the orientation of a small axisymmetric particle
 settling in a turbulent flow. In Refs.~\cite{Siew14a,Siew14b,Gus17,Jucha2018,Naso2018}
 this effect was neglected. Here we considered a limit of the problem where it is dominant,
 but where turbulent fluctuations still matter. Our goal was to compute the distribution of orientations
 of a spheroid in turbulence, to work out how the torques due to convective fluid inertia and due
 to the turbulent velocity gradients affect the orientation distribution.
 In general the angular dynamics of the settling particle is very complicated. Here we looked at
 a limit in which the problem becomes tractable: we assumed small Stokes number (a dimensionless
 measure of particle inertia) and  large settling number (dimensionless settling speed).
For small Stokes numbers the dynamics is overdamped. For large values of the settling number,
the angular dynamics becomes persistent: it relaxes much more rapidly than the fluid-velocity gradients change.
In this limit  the angular dynamics follows the fixed points determined
by the instantaneous fluid-velocity gradients, and our theory for the orientation distribution  relates the shape of the distribution to that of the instantaneous fluid-velocity gradients encountered by the settling particle. Our predictions are in excellent agreement with numerical statistical-model simulations, and with simulations using KS turbulence at large $\sv$ and small enough $\st$.

At large $\sv$ the orientation distribution is very narrowly centered around
the orientation the settling particle would assume in a quiescent fluid, in the absence of flow.
The overdamped theory predicts that the variance of the distribution is proportional to $\sv^{-4}$ for
 large $\sv$, and it determines how the
 prefactor depends on aspect ratio $\lambda$ of the particle. In  the limit $\lambda\to\infty$ the variance was computed
in Ref.~\cite{Kramel}.

 We demonstrated that the overdamped theory  breaks down at finite
Stokes numbers, when the settling number exceeds a threshold determined by $\st$.
In this regime particle inertia matters. Klett \cite{Kle95} proposed a theory for the orientation variance
for nearly spherical particles, taking into account particle inertia in the angular dynamics.
His theory assumes that this dynamics is driven by the fluid-velocity gradients experienced by the settling particle,
and that these gradients are uncorrelated in time so that diffusion approximations can be applied.
Klett's theory predicts that the variance
is proportional to $\sv^{-2}$, and we do observe this scaling for very large $\sv$, so large that the  settling time is the smallest time scale of the inertial dynamics. But to derive a theory
from first principles it is necessary to take into  account particle inertia not only in the angular dynamics
but also in the centre-of-mass motion, resulting in additional fluctuating terms in the angular equation of motion that are
expected to change the orientation variance.
More importantly, our simulations also show that particle inertia gives rise to a complex dependence of the orientation
variance on particle shape, on the Stokes number, and upon the settling number.
When the  variance is 
small, it may be possible to derive a theory for the variance using small-angle approximations. But this
remains a question for the future.
 
Here we applied our theory only to prolate particles. It is of interest to consider oblate particles too, 
because flat disks and slender rods have qualitatively different shape factors (Fig.~\ref{fig:shape_factor}). 
We therefore expect that the effect of particle inertia on the angular dynamics of flat disks can be quite different from that on slender rods.
Also, we considered only the leading order in the inverse settling number, but the overdamped theory allows us to take into account higher-order corrections in this parameter. Such corrections
 change the relation
between the fixed points of the angular dynamics and the fluid-velocity gradients experienced
by the particle. This modifies
the form of the distribution of $n_g$, and it may explain the overshooting seen in Fig.~\ref{fig:variance}(b)
at moderate values of $\sv$, but the details remain to be worked out.

Here we analysed a limit of the problem where the fluid-inertia torque dominates the angular motion.
In Refs.~\cite{Siew14a,Siew14b,Gus17,Jucha2018,Naso2018}, by contrast, this torque was neglected. The question
is thus whether
one  can  find regions where inertial torque does not dominate. This is considered in Ref.~\cite{Sha19}. The simulations
described there show that the fluid-inertia torque can be smaller than Jeffery's torque only  when $\re_\lambda$  is small.
In a very turbulent flow, when  $\re_\lambda$ is large,  the torque induced by fluid inertia  is always dominant. More precisely,
when  the ratio of the correlation length over the Kolmogorov length is large, $\ell/\eta_K \propto {\re_\lambda}^{1/2} \gg 1$, then the only possible orientation bias corresponds to non-spherical particles settling with their broad sides down, the limit considered here.

The experiments measuring the orientations of rods settling in a vortex flow described in
Ref.~\cite{Lop17} are performed in the overdamped limit.
In the future we intend to apply the theory outlined in Section \ref{sec:advective} to spheroids settling
in a two-dimensional vortex flow, using the fact that   the fixed points of the angular dynamics can be found
explicitly as functions of the fluid-velocity gradients in two spatial dimensions. We will analyse
the effect of particle shape by considering the angular dynamics of flat disks settling in such flows.
Figure \ref{fig:shape_factor} indicates that the behaviour could be quite different from that of rods,
because the shape factors are so different. This two-dimensional system is well suited to study the effects
of finite Stokes numbers in more detail, because the two-dimensional dynamics is much simpler
than the three-dimensional turbulent dynamics.

The overdamped theory [Eq.~(\ref{eq:var_general})] assumes that $\sv$ is large,
and that $\st$ is small enough. Since $\sv = \st \,{g\tauK^2}/{\etaK} = \st/{\rm Fr}$,
this requires some discussion. Here  ${\rm Fr} = \etaK/(g\tauK^2)$ is the Froude number \cite{Dev12}.
We conclude that the Froude number must be small for the overdamped theory to work quantitatively. In turbulence
${\rm Fr} \sim \mathscr{E}^{3/4}/(g\nu^{1/4})$ where $\mathscr{E}$ is the dissipation rate per
unit mass. Using $\nu \sim 10^{-5}\,{\rm m}^2 {\rm s}^{-1}$ and
$g = 10\, {\rm m} s^{-2}$ we find that ${\rm Fr}$ ranges from $0.002$ at $\mathscr{E}=1 \,{\rm cm}^2 {\rm s}^{-3}$ to $0.3$ at $\mathscr{E}=1000\, {\rm cm}^2 {\rm s}^{-3}$. So we require modest values
of the dissipation rate per unit mass, $\mathscr{E}$, for the theory to work quantitatively. This is the limit
where gravity dominates over the turbulent fluctuations, the limit we intended to describe.

In the future it is necessary to address possible shortcomings of our model which approximates the inertial contributions to force and torque by those for a homogeneous steady flow.  Even in the steady case it remains an open question how to model
the torque when $\re_p$ and $\sqrt{\re_s}$ are of the same order, even if both dimensionless numbers  are small. Furthermore, turbulent flow is unsteady. While it is common practice to
use steady approximations for the instantaneous force and torque (as we do here) it is not known
how to compute contributions to the torque due to unsteadiness
for general inhomogeneous flows. We expect that the methods presented in Ref.~\cite{Candelier2018} can
be generalised to treat at least spatially linear, unsteady flows. Finally, to justify our model for the inertial torque it is necessary that $\re_p$ is small. At the same time we assumed that $\sv$ is large.
From the definitions (\ref{eq:re_defs}) and (\ref{eq:dimless_parameters})  of these dimensionless numbers we see that
$\rep = ({a}/{\etaK})({\sv}/{A_\perp})$.
To satisfy both requirements we must therefore assume the particles to be much smaller than the Kolmogorov length.
Since $\etaK \sim(\nu^3/\mathscr{E})^{1/4}$ this condition is more easily met when $\mathscr{E}$ is small.
In the slender-body limit, Khayat \& Cox \cite{Kha89} obtained an improved approximation
for the inertial torque, valid
for larger $\re_p$, which was tested in Ref.~\cite{Lop17} and was found to agree better
with the experiments at larger $\re_p$. But corresponding corrections for other particle shapes are not yet known.

\ack
BM and AP thank E. Guazzelli for enlightening discussions.
 KG and BM were supported by the grant {\em Bottlenecks for particle growth in turbulent aerosols} from the Knut and Alice Wallenberg Foundation, Dnr. KAW 2014.0048, and in part by VR grant no. 2017-3865.
 AP and AN acknowledge support from the IDEXLYON project (Contract ANR-16-IDEX-0005)
under University of Lyon auspices.  Computational resources were provided by C3SE and SNIC, and PSMN.
\vspace*{5mm}

%\bibliographystyle{unsrt}
%\bibliography{dbib}

\end{document}